# Towards Enhanced Classification of Abnormal Lung sound in Multi-breath: A Light Weight Multi-label and Multi-head Attention Classification Method

Yi-Wei Chua[1]*, Yun-Chien Cheng[1]*,

[1] Department of Mechanical Engineering, College of Engineering, National Yang Ming Chiao Tung University, Hsin-Chu, Taiwan

*Corresponding author: yccheng@nycu.edu.tw, Yiweicc0306@gmail.com

# Abstract

This study aims to develop an auxiliary diagnostic system for classifying abnormal lung respiratory sounds, enhancing the accuracy of automatic abnormal breath sound classification through an innovative multi-label learning approach and multi-head attention mechanism. Addressing the issue of class imbalance and lack of diversity in existing respiratory sound datasets, our study employs a lightweight and highly accurate model, using a two-dimensional label set to represent multiple respiratory sound characteristics. Our method achieved a 59.2% ICBHI score in the four-category task on the ICBHI2017 dataset, demonstrating its advantages in terms of lightweight and high accuracy. This study not only improves the accuracy of automatic diagnosis of lung respiratory sound abnormalities but also opens new possibilities for clinical applications.

## 1. Introduction

In recent years, the harm of influenza to humans has been increasing, and the rapid spread of COVID-19 disease has exacerbated this issue, leading to the death of most patients due to respiratory system abnormalities. Before this epidemic outbreak, three respiratory diseases had already become one of the top ten causes of death globally[1]. According to the World Health Organization's report, respiratory diseases have become one of the main causes of death in society, including the "big five" respiratory diseases: asthma, chronic obstructive pulmonary disease (COPD), acute lower respiratory infections, lung cancer, and tuberculosis. COPD ranks third among global fatal diseases, claiming the lives of 3.2 million people annually, accounting for an astonishing 81.7% of all chronic respiratory disease deaths[2] .

As pulmonary diseases have garnered increasing attention, early diagnosis of these diseases has also become a focal point of concern. Generally, monitoring respiratory sounds through auscultation is the traditional method for assessing a patient's respiratory health; for this purpose, doctors commonly use stethoscopes as a clinical tool for diagnosing pulmonary diseases and abnormalities. The main aim of a stethoscope is to identify variations in respiratory sounds within a given time frame, such as Wheezes, Crackles, and Stridor. Crackles are brief, explosive, non-musical sounds, often occurring in patients with substantive lung diseases, such as pneumonia, interstitial pulmonary fibrosis (PF), and pulmonary edema [3]. Wheezes are abnormal breathing sounds associated with airway diseases such as asthma and Chronic Obstructive Pulmonary Disease (COPD), characterized by a high-pitched tone, lasting over 80 milliseconds [4]. Wheezing is described as a continuous whistling and hissing sound, superimposed on normal breathing. Wheezes are caused by airway narrowing, leading to restricted airflow [5]. Stridor is a continuous airway sound similar to wheezing, characterized by a hissing and musical quality. This sound is mainly heard during inhalation, but may sometimes appear during exhalation or in both phases. Unlike wheezing, stridor is caused by airflow turbulence in the throat or bronchial tree and is usually associated with upper respiratory tract obstruction [6]. These different types of respiratory sounds play a crucial role in diagnosing various lung and airway diseases.

Using a stethoscope to listen to lung sounds is a traditional technique and the most popular diagnostic method among specialists for the preliminary assessment of respiratory diseases. The advantages of auscultation include being a non-invasive diagnostic method and an effective auxiliary diagnostic tool, aiding in the diagnosis and differentiation of various respiratory diseases. However, there are limitations to this diagnostic approach, as described below:

(1) Due to varying interpretations of respiratory sounds by different medical professionals, there can be subjectivity in the diagnosis. This subjectivity may affect the accuracy and consistency of diagnoses, thereby complicating the determination of treatment plans for patients [7].

(2) The similarity of various abnormal respiratory sounds can cause diagnostic confusion, such as the wheezing of bronchial asthma and the stridor of Vocal Cord Dysfunction (VCD), which are often confused during the preliminary diagnosis of airway obstruction during physical activities[8]. Both sounds are described as continuous, high-pitched, musical sounds with sinusoidal waveforms, exhibiting periodicity in the time domain. This similarity can lead to confusion in the initial diagnostic phase,

(3) Necessitating more precise diagnostic tools or methods to correctly differentiate them for more effective treatment planning.

In summary, quantifying the analysis of recorded lung breathing sounds can provide a systematic approach to diagnosing different respiratory conditions by automatically classifying acoustic patterns. This automation not only helps reduce subjectivity in the diagnostic process by physicians but also provides faster and more accurate results, especially in situations where rapid assessment of a large number of patients is needed, such as during epidemics or emergencies. Additionally, the advantages of automation can be used for long-term observation of the effects of medication post-treatment and optimizing individual treatment plans.

For the recognition of abnormal respiratory sounds, common data feature transformations include one-dimensional time-domain audio raw data and two-dimensional data in the time-frequency joint domain, such as short-time Fourier transform spectrograms, Mel-Frequency Cepstral Coefficients (MFCCs), and Constant-Q Transform (CQT)[9]. Although one-dimensional raw audio signals require less computation and do not need amplitude scaling like frequency domain signals, their limitation lies in the need for a large amount of data to achieve results similar to two-dimensional spectrogram [10]. Given the diversity and noise present in real-world respiratory sounds, time-domain audio features may not effectively distinguish between signal and noise, impacting the performance of model training. Therefore, using frequency transformation and spectrograms to obtain more signal features, including spatial and frequency energy characteristics, can enhance the generalization ability of machine learning models. Studies, such as those by Alice al, have confirmed that two-dimensional data representations are more effective in understanding and analyzing pulmonary abnormalities compared to one-dimensional data [11]. Although existing machine learning (ML) based sound recognition systems generally use manually designed time-frequency joint domain feature transformations and achieve relatively good performance, these transformations are still limited by fixed feature biases[12]. For instance, while fixed Mel scale and logarithmic compression are commonly effective, we cannot guarantee that they always provide the best performance for downstream tasks. Even though these biases are beneficial for matching human perception in areas like speech recognition or music understanding, they might not be advantageous in domains where mimicking human hearing is not critical. Therefore, to achieve optimal performance, common practices include replacing Fourier transformations with learnable filters and customizing different learnable or specifically constrained filter sets, along with various compression and normalization methods, tailored to specific tasks.

In recent years, with the advancement of machine learning technology, many identification algorithms such as logistic regression and Gaussian mixture models have been widely used in diagnostics and telemedicine, especially in predicting respiratory diseases. For example, Palaniappan's team used MFCC features combined with one-way ANOVA and classifiers like Support Vector Machine (SVM) and K-Nearest Neighbors (KNN) to categorize respiratory sounds, finding that KNN outperformed SVM in differentiating pathological and normal lung sounds[13]. However, traditional machine learning methods have limitations, requiring extensive experimentation to achieve the best combinations. Current research trends toward using deep learning, such as RNNs and CNNs, to extract higher-level semantic features. These methods are more suitable for handling long sequence data and attention mechanisms, improving model performance on specific tasks. For instance, a CNN-based model proposed by Siddhartha al, integrating device-specific fine-tuning and data augmentation techniques, achieved higher classification accuracy on the ICBHI dataset[14]. Additionally, Bae's team demonstrated the application of pre-trained models and innovative augmentation techniques, significantly improving performance on the ICBHI dataset[15]. Other studies [16-19] also show that combining different training methods and architectures can enhance model performance. Overall, for diagnosing pulmonary respiratory sounds, the design of deep learning algorithms needs to consider the model's lightweight and effectiveness to be integrated with devices like electronic stethoscopes.

In the field of respiratory sound recognition, there are currently three main public respiratory sound databases: the 2017 International Conference on Biomedical and Health Informatics (ICBHI) database, the Pediatric Respiratory Sound database (SPRSound), and the King Abdullah University Hospital (KAUH) database. The age range of participants in these databases is broad, and there are variations in the labeling annotations of the respiratory sounds, annotated by different doctors without a unified standard. Additionally, the distribution of respiratory sound participants in each database is uneven, and some databases have insufficient data, as shown in Figure 1.

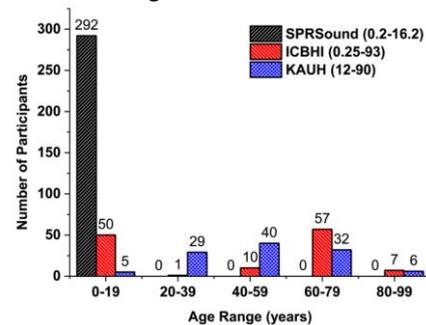

Figure 1: The number of patient's distribution in different dataset

In the classified 4-classes task of the ICBHI dataset, conflicts were found among category labels such as normal, crackle, wheezes, and crackle&wheezes. For instance, in the crackle&wheezes category, the sound signals contain features of both crackle and wheeze. This ambiguous classification could cause blurred classification boundaries in deep learning model training. Additionally, the imbalance of data categories and the lack of data diversity may reduce the robustness of the model, potentially leading to convergence confusion during training, thus impacting performance.

In recent years, the use of pure attention mechanism architectures like the transformer has significantly improved performance in many downstream tasks. In the field of respiratory sound detection, many studies adopting the Audio Spectrogram Transformer pre-trained on the Audioset architecture have surpassed state-of-the-art performance[20]. However, designing lightweight models with high accuracy for deployment on end devices is closer to practical application. Therefore, this study's contributions to practically applicable architectural designs are as follows:

(1) This study will design a lightweight model architecture and incorporate attention mechanisms in the classifier to achieve performance comparable to the best current models for classifying lung respiratory sound anomalies.
(2) This study employs a multi-label approach to replace the original category labels, aiming to reduce convergence confusion in model training.
(3) The study explores the impact of a learnable spectral front-end module on model performance."

## 2. Related Work

### 2.1. Learnable spectral front-end module

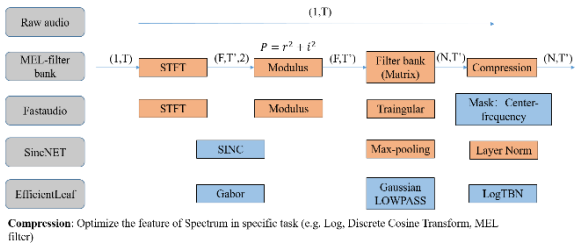

Figure 2: The learnable spectrum modulus describe in different method

Sainath al initially attempted to learn filters based on mel-filterbanks, with the filterbank being initialized on the mel scale and then learned along with other network components, using spectrograms as input [20]. Contrarily, Sainath al[10] and Hoshen al[21] later proposed directly learning convolutional filters from raw waveforms, initializing them with Gammatone filters [22]. SincNET, based on a parametric Sinc function, implements learnable band-pass filters. Unlike standard CNNs, SincNET learns only the low and high cutoff frequencies directly from the data, rather than learning all elements of each filter[23]. EfficientLeaf, an improvement based on Leaf, uses Gabor filters, which are generated by combining sinusoidal signals with Gaussian kernels [24]. These filters offer ideal characteristics, achieving the best balance in time and frequency localization, making them an excellent choice for convolutional networks containing filters of finite size.

The second step of the learnable front-end is to downsample the output of the filter bank to a lower sampling rate. Past studies have explored the potential applications of learnable filter banks, such as the higher degree of freedom learnable filter banks implemented by nnAudio[25]. However, T. Sainath al[26] pointed out that their improvements in learning are limited. Additionally, Fu's team conducted detailed studies on the effects of shape and positivity constraints on filter banks [27]. In the Efficientleaf method [28], the authors expanded Zeghidour et al.'s pooling layer in two ways [29], including using independent low-pass filters for each input channel, and parameterizing these low-pass filters to achieve Gaussian pulse responses. Finally, per-channel energy normalization (PCEN) involving sequential computation was replaced with learnable logarithmic compression, temporal median subtraction, and Temporal Batch Normalization (LogTBN), all of which are parallelizable and thus faster to compute on graphic processing units.

### 2.2. Deep Learning Architecture

One major challenge in the task of respiratory sound classification is the limited number of samples required to train large networks, compounded by the imbalance in category numbers within the currently available public datasets. To address this issue, past research often utilized traditional visual models pre-trained on ImageNet (IN) or AudioSet (AS) to compensate for the lack of training samples. AST, a model entirely based on the attention mechanism, made significant progress in several audio processing tasks after being sequentially pre-trained on ImageNet and AudioSet[20]. Although transformers perform excellently, they are resource-intensive in terms of memory and computation and challenging to deploy on devices. PSLA introduced a series of training techniques such as ImageNet pre-training, balanced sampling, label enhancement, and model aggregation, achieving the highest mean average precision (mAP) of .5671 on FSD50K using a combined model [30]. PSLA utilized multiple architectures based on EfficientNet (B0, B2 with attention), with the base model EfficientNet-B0 combining single-head attention methods, demonstrating that the selection of training techniques could achieve good performance even with smaller models on the AudioSet dataset. Lu et al. designed the CAB attention mechanism architecture combined with classifiers, effectively reducing the classifier's parameters and complexity, resulting in more stable training and better performance[31]. In the field of respiratory sound detection, Moummad and others used a CNN6 model pre-trained on AudioSet, combined with a contrastive learning architecture, achieving results comparable to advanced methods. Therefore, we will also attempt to select different lightweight model architectures, pre-train them on audio domain datasets, and combine attention mechanisms in classifiers to achieve high accuracy and lightweight objectives in downstream respiratory sound classification tasks.

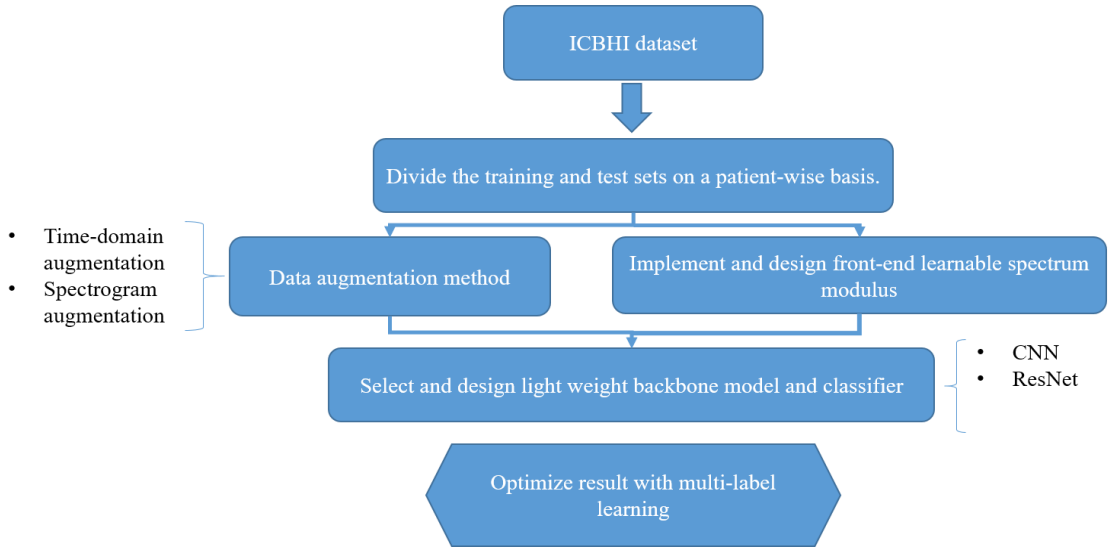

Figure 3: Experiment process

## 3. Material and Methods

### 3.1. ICBHI 2017 Dataset

This study utilized the respiratory sound database organized at the International Conference on Biomedical Health Informatics (ICBHI 2017)[32]. The ICBHI dataset comprises 6,898 respiratory cycles, totaling approximately 5.5 hours, and is officially divided into a training set (60%) and a test set (40%). Notably, in the split of the training and test sets, the data of patients do not overlap between the two. Each respiratory cycle is labeled as one of four categories: Normal, Crackle, Wheeze, or Crackle & Wheeze. The training set contains a total of 539 recordings from 79 patients, including 1,215 crackle cycles, 501 wheeze cycles, 363 crackle & wheeze cycles, and 2,063 normal respiratory cycles. Similarly, the test set includes 381 recordings from 49 patients, with a total of 649 crackle cycles, 385 wheeze cycles, 143 crackle & wheeze cycles, and 1,579 normal respiratory cycles.

### 3.2. Experimental Procedure

The experimental process of this study is illustrated in Figure 7. Initially, data collection will be conducted, with patient-based division into training, validation, and test sets. Subsequently, a combination of learnable front-end architectures or direct transformation into feature spectrograms will be utilized, along with data augmentation methods. In terms of model design, the focus will be on selecting lightweight models for use and training weights. Then, adjustments will be made to the classifier's architecture to optimize performance, followed by the integration of multi-label methods for further optimization. Finally, the study will conduct multiple tests, including:

(1) Comparative testing of various models' results.

(2) Validation of the performance improvements due to pre-training and its dataset sources.

(3) Exploration and integration of attention mechanisms in the classifier's architectural design.

(4) Verification of the effectiveness of multi-label methods on the dataset used in our study.

### 3.3. Data Preprocessing

The sampling rate of audio recordings ranges from 4 kHz of the ICBHI to 44.1 kHz, and we resample all recordings to 16 kHz. As each patient's respiratory cycle duration varies from approximately 0.2 seconds to 16.2 seconds, with an average of 2.7 seconds, we employ circular padding to extend the duration to 8 seconds for consistency in the model's input spectrogram or raw audio size. If using a learnable spectrogram front-end module, we randomly alter the speed and pitch of the audio for data augmentation. Otherwise, we convert the audio signal into a time-frequency representation Mel spectrogram. This is done using 64 Mel filters, a window size of 1024, a hop size of 512, a minimum frequency of 50 Hz, and a maximum frequency of 2000 Hz, as both wheezes and crackles fall within this frequency range [33]. The resulting spectrogram size is (64×256). Subsequently, we apply partial masking data augmentation on the time or frequency axis.

### 3.4. Model Overview

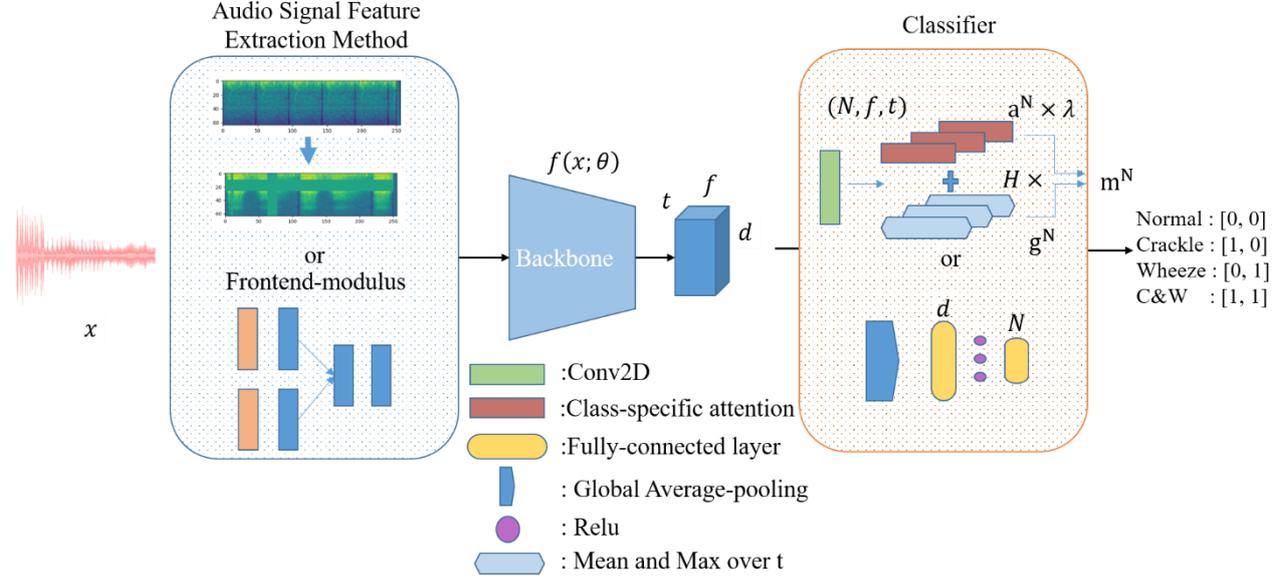

Figure 4: Multi-breath model architecture

The model designed in this study is named Multi-breath, as shown in Figure 3. The feature transformation of audio signals will attempt to use either Mel-spectrogram or spectrogram information obtained from a learnable front-end architecture as input for the deep learning structure. When selecting different models, the goal will be based on lightweight design, subsequently outputting d features of size (f,t). Here, f represents the size of the frequency feature dimension, and t represents the size of the time feature dimension. The other method was discussed in following small section.

*3.4.1. Multi-label Learning method*

Traditional supervised learning is one of the most extensively studied paradigms in machine learning, where each object (case) in the real world is represented by a single instance (feature vector) and associated with a single label. Specifically, assuming Y represents the label space, the goal of traditional supervised learning is to learn a function from the training set { ⊢ (x_i,y_i )⊣ |1≤i≤m}. Here, x_i ϵX is an instance describing object attributes (features), and y_i ϵY is the corresponding label describing its semantics{Zhang, 2013 #51}. Thus, a fundamental assumption adopted in traditional supervised learning is that each instance belongs to only one concept, i.e., it has a unique semantic meaning. However, in many learning tasks, the above simplifying assumption is not applicable, as objects in the real world can be complex and possess multiple semantic meanings simultaneously. For example, in a symphony, there exist many different pieces of information, such as piano sounds, classical music, and Mozart. Therefore, the multi-label approach considers that an object may have multiple semantic meanings. A straightforward solution is to assign a set of appropriate labels to the object to explicitly express its semantics. In the task of classifying the lung respiratory dataset, we believe that categorizing Crackle, Wheeze, and Crackle & Wheeze into three separate classes could cause confusion in model training, as the Crackle & Wheeze category might contain features of the other two classes. Therefore, we attempt to use a multi-label approach for task learning.

The task of multi-label learning is to learn a function f: X → Y to predict the appropriate set of labels for unseen instances. In this task, each instance is associated with a set of category labels, represented by a sparse binary label vector. A value of 1 indicates that the instance belongs to the category at that vector position, and 0 indicates non-belonging. Thus, $y_j \in Y = \{0,1\}^m$, where m denotes the total number of labels in the set, in other words, it represents the number of label categories each instance might be associated with in a multi-label learning task. Given an unseen instance x ∈ X, the learned multi-label classification function f(·) outputs f(x) ∈ Y, where the output vector is the predicted score for whether it belongs to m labels, then passed through a sigmoid activation function σ to bring the prediction scores between 0 and 1, which is:

$$\sigma(f(x)) = \frac{1}{1 + e^{-f(x)}}$$

Equation 1

For each element in the output vector, if the prediction score is greater than or equal to a threshold τ (set to 0.5), then the label is predicted to be present (1); otherwise, it is predicted to be absent (0). Mathematically, this can be expressed as:

If $\sigma(f(x))_i \geq \tau$, then $y_i = 1$; otherwise, $y_i = 0$, where i = 1, 2, ..., m represents the index of the labels.

*3.4.2. Implement multi-head attention in classifier*

Global average pooling and max pooling are two commonly used techniques in deep learning for feature extraction. Global average pooling calculates the average value of each channel of the feature map, while max pooling extracts the maximum value of each channel. These two methods capture different features of the data. Inspired by these methods, our classifier aims to design a scientific attention mechanism to replace the fully connected layer, thereby reducing model parameters and increasing accuracy.

Suppose we are given a feature spectrogram, first processed by a feature extractor $f(x:\theta)$ to obtain a feature vector $x \in R^{d \times f \times t}$, where d, f, t represent feature dimensions, frequency dimensions, and time dimensions, respectively. In our study, we used a 64×256 feature spectrogram, assuming that the feature map obtained through the feature extractor is 512×4×16. The input to the classifier can be decoupled into $x_1, x_2, \ldots, x_{64} (x_i \in R^{512})$, and then passed through a one-dimensional convolution layer (all are fully connected layers) as the classifier. For the i-th category, the classifier's parameters are $C_i \in R^{512}$, with N being the number of categories. When using different settings, constants like 64 can be adjusted accordingly. We refer to the CSRA architecture to improve the multi-head class-specific attention[34]. We now define the class-specific attention score for the i-th class and the j-th position as:

$$s_j^i = \frac{\exp(Tx_j^T C_i)}{\sum_{k=1}^{64} \exp(Tx_k^T C_i)}$$

Equation 2

The formula $\sum_{j=1}^{64} s_j^i = 1$ ensures that for any specific position j, the sum of the probabilities of all categories i is 1, conforming to a probability distribution. In the context of the softmax function, the temperature parameter T is used to control the 'sharpness' or 'smoothness' of the output probability distribution. When T is larger, the probability distribution is smoother, and the probability differences between categories are smaller. When T is smaller, the probability distribution is sharper, and the maximum probability becomes more pronounced. Then, we can define the class-specific feature vector for class i as a weighted combination of the feature tensor, where the attention scores $s_k^i$ ($1 \leq k \leq 49$) of class i are the weights, as follows:

$$a^i = \sum_{k=1}^{64} s_k^i x_k$$

Equation 3

For the global feature spectrogram, we first perform global averaging across the time dimension t, then apply max pooling and average pooling on the resulting mean (N, f) in the f dimension, and finally add up the values to obtain $g^i$, which is as follows:

$$GAP_t = \frac{1}{N \times F} \sum_{n=1}^{N} \sum_{f=1}^{F} R_{n,f,t}$$

Equation 4

$$GMP_f = \max_{1 \leq n \leq N} R_{n,f}$$

Equation 5

$$GAP_f = \frac{1}{F} \sum_{f=1}^{F} R_{n,f}$$

Equation 6

$$g^i = GAP_f + GMP_f$$

Equation 7

We use $g^i$ as the primary feature, which is obtained by multiplying by $\lambda$ and the class-specific attention mechanism $a^i$ to get $f^i$, with the formula as follows:

$$f^i = g^i + \lambda a^i$$

Equation 8

The temperature hyperparameter *T* can be challenging to adjust, and different categories may require different temperatures (or different attention scores). Therefore, we referred to CSRA and further designed a multi-head attention extension, where each branch utilizes different temperatures *T* but shares the same λ. We denote the number of attention heads as *H*. To avoid adjusting the temperature *T*, we either choose a single head *(H = 1)* and fix the temperature *T = 1*, or use multi-head attention *(H > 1)* and fix the temperature sequence $T_1$, $T_2$, ..., $T_H$ In addition to *H = 1*, we also used *H = 2, 4, 6*. Specifically:

- When H = 2, $T_1$ = 1 and $T_2 = \infty$ (eg. Max pooling)

- When H = 4, $T_{1:3}$ = 1, 2, 3 and $T_4 = \infty$;

- When H = 6, $T_{1:5}$ = 1, 2, 3, 4, 5 and $T_6 = \infty$;

**3.5. Equipment**

This study utilized a server with ASUS Z790-A GAMING WIFI 6E, powered by Intel I9-13900K, and equipped with MSI RTX 4090 GAMING X TRIO 24G.

### 3.6. Implement detail

To prevent overfitting when using feature spectrograms, we employ time and frequency masking augmentation methods, with a maximum coverage of 20 frames in the time domain and 40 bins in the frequency domain. In this study, while training with Multi-label, we used the Adam optimizer with a learning rate of 1e-3, cosine annealing schedule, and a batch size of 64. The classification loss function used is binary cross entropy (BCE Loss), which first converts the model's raw output into probability values via the Sigmoid function and then calculates the binary cross-entropy loss between predicted values and actual labels. When training a single category, cross entropy loss is used.

## 4. Result

### 4.1. Evaluation Methods

For all deep learning architectures, this study will will evaluate the classification performance based on the assessment metrics of the ICBHI 2017 challenge. We assess using the Score, which is the average of Specificity (Sp) and Sensitivity (Se). The definitions of specificity and sensitivity are as follows:

$$s_p = \frac{C_n}{N_n}$$

Equation 9

$$S_c = \frac{S_e + S_p}{2}$$

Equation 10

$$s_e = \frac{C_c + C_w + C_{c\&w}}{N_c + N_w + N_{c\&w}}$$

Equation 11

Where $C_i$ and $N_i$ are, respectively, the number of correctly classified samples and the total number of samples in category i ∈ {Normal, Crackle, Wheeze, Crackle & Wheeze}. To simplify and align with the evaluation metrics of the ICBHI dataset, we use the trained classifier for four-category classification, followed by calculating the two-category version of Sensitivity (Se).

### 4.2. Comparison of different models

This section aims to explore the comparison of different lightweight models for classifying abnormal respiratory sounds. Table 1 shows the results of different models on the test set, including a larger parameter model CNN14 as the upper bound for comparison, followed by ResNet22, CNN6, and MobileNetV2 in descending order of model parameter size. All model architectures used pre-trained weights on AudioSet as the initial point for downstream tasks. The results show that the larger CNN14 has the highest score, however, when testing the MobileNetV2 architecture, the results significantly drop by about 8% in score. The CNN6 architecture, on the other hand, aligns with our research goal of having low parameters and excellent performance. This study will take CNN as the primary lightweight architecture for further improvements.

Table 1. Comparison of different models

| Model | Parameter (M) | Sp | Se | Score |
|---|---|---|---|---|
| CNN14 | 76.53 | 0.8 | 0.385 | 0.594 |
| ResNet22 | 11.27 | 0.654 | 0.4 | 0.5279 |
| MobileNetV2 | 2.88 | 0.7036 | 0.3347 | 0.519 |
| CNN6 | 4.37 | 0.6846 | 0.424 | 0.5546 |

### 4.3. Comparison of multi-label learning method in different models

This section aims to discuss the performance of using multi-label learning methods. In Table 2, we find that the combination of multi-label training methods with CNN14 does not show significant improvement. It is speculated that larger models have better semantic ability to extract features from spectrograms, preventing accuracy decline due to category ambiguity. However, in different smaller model architectures, multi-label methods have achieved significant improvements, with the combination with CNN6 architecture showing the most improvement, up to 2.8% in score. This proves that alleviating the interference of category noise in training data is a viable direction. It was also observed during the training process that the testing loss was relatively lower.

Table 2. Multi-label with different models

| Model | Parameter (M) | Sp | Se | Score |
|---|---|---|---|---|
| CNN14 | 76.53 | 0.777 | 0.41 | 0.594 |
| ResNet22 | 11.27 | 0.633 | 0.441 | 0.536 (+0.008) |
| MobileNetV2 | 2.88 | 0.789 | 0.26 | 0.5245 (+0.005) |
| CNN6 | 4.37 | 0.722 | 0.448 | 0.583 (+0.028) |

### 4.4. Comparison of classifier implement with multi-head attention in different models

This section aims to discuss the performance of the combination of multi-head class-specific attention mechanisms with classifiers in different models. As seen in table 3, most models that incorporated this method had a positive impact, not only enhancing model performance but also replacing fully connected layers, thereby reducing model parameters. However, there was a decrease in accuracy with the MobileNetV2 model. We speculate that this may be due to the lower feature dimension output after feature extraction, preventing the attention mechanism from capturing enough feature information to grasp the spatial relations and important features on the spectrogram in time and dimensions. The use of multi-head attention mechanisms adjusts the focus range of each attention head by tuning different temperature parameters. Through this method, the model can find a balance between fine-grained and coarse-grained features, thus enhancing its overall representational ability.

Table 3. Multi label with multi-head attention classifier in different backbone

| Model | Parameter (M) | Sp | Se | Score |
|---|---|---|---|---|
| CNN14 | 75.53 (-1) | 0.7809 | 0.42 | 0.6 (+0.006) |
| ResNet22 | 11.22 (-0.05) | 0.6852 | 0.4027 | 0.544 (+0.008) |
| MobileNetV2 | 2.25 (-0.03) | 0.682 | 0.3246 | 0.503 (-0.02) |
| CNN6 | 4.32 (-0.05) | 0.73 | 0.4537 | 0.592 (+0.009) |

**4.5. Comparison of learnable spectrum front-end modulus with our design models.**

This chapter primarily focuses on whether the combination of traditional Fourier spectrogram features and learnable front-end spectrogram features can enhance performance for this task. According to the team's research, having a sufficient dataset allows the learnable front-end to effectively learn diverse data and adjust its parameters to extract key feature representations and optimize the model's feature extraction[Poirè, 2022 #58]. Therefore, this study also attempts to adjust the training and test set ratios from 60:40 to 80:20 to compare the architecture combined with learnable spectrogram features. For the learnable front-end architecture, we used Efficientleaf for testing. The results show that CNN14, when combined with the front-end architecture, does not improve accuracy despite an increase in the training dataset size. However, when combined with the CNN6 model, which has fewer parameters, the results with an 80:20 split of the training dataset are similar to those of traditional Fourier spectrogram transformations. We speculate that a balance between the complexity of the model and its combination with the front-end architecture is essential for improving accuracy. Future designs might similarly pre-train the front-end architecture or experiment with different methods in the front-end architecture.

Table 4. Multi label with our design models

| Frontend Modulus | Model | Sp | Se | Score |
|---|---|---|---|---|
| 60/40 Training, Testing split | | | | |
| | CNN14 | 0.7809 | 0.42 | 0.6 |
| EfficientLeaf | CNN14 | 0.8075 | 0.32 | 0.564 (-0.036) |
| | CNN6 | 0.73 | 0.4536 | **0.592** |
| EfficientLeaf | CNN6 | 0.782 | 0.371 | 0.576 (-0.016) |
| 80/20 Training, Testing split | | | | |
| | CNN14 | 0.8781 | 0.7097 | 0.7939 |
| EfficientLeaf | CNN14 | 0.893 | 0.6292 | 0.7613 (-0.0326) |
| | CNN6 | 0.83518 | 0.6945 | 0.7649 |
| EfficientLeaf | CNN6 | 0.864 | 0.67 | 0.767 (+0.002) |

**4.6. Comparison between state-of-the art methods.**

This chapter describes the overall respiratory sound classification performance of our study compared to other methods on the ICBHI dataset. In the comparative results in table 5, we also explored the model's performance on different pre-trained datasets, including ImageNet (IN) and AudioSet (AS). Bae's team's use of different datasets for pre-training on transformers shows that AS, with data styles more similar to our target task, achieves better accuracy, and combining the two pre-training datasets also shows significant improvement. Consequently, in our study, the initial parameters of the model were pre-trained using AS. Our results show that the CNN14 model architecture combined with our proposed multi-label multi-head attention classifier architecture outperforms methods proposed by other studies and is comparable to the transformer architecture pre-trained on AS and IN [Bae, 2023 #37]. In terms of selecting a lightweight model target, we compared it with the method proposed by Moummad's team, which also uses a lightweight model [Moummad, 2023 #38]. Their SCL combined with CNN6 primarily utilized a supervised contrastive learning architecture, replacing the cross-entropy training method, and integrated multiple datasets for the model to learn more helpful feature information. In comparisons using CNN6 as the model architecture, our method showed significant improvement in evaluation metrics compared to the other two methods, approximately a 1.2% score increase.

Table 5. Multi label with multi-head attention classifier in different backbone

| Model | Pretrain | Sp | Se | Score |
|---|---|---|---|---|
| RestNet34 | IN | 0.714 | 0.39 | 0.552 |
| ResNet50 | IN | 0.763 | 0.374 | 0.569 |
| ResNet50(Co-tuning) | IN | **0.7934** | 0.372 | 0.583 |
| AST | IN | 0.787 | 0.388 | 0.587 |
| AST | IN+AS | 0.77 | 0.42 | 0.596 |
| MixCL+AST | IN+AS | 0.82 | **0.431** | **0.624** |
| CNN14 [ours] | AS | 0.7809 | 0.42 | 0.6 |
| Light weight model | | | | |
| SCL-CNN6 | AS | 0.76 | 0.39 | 0.57 |
| Meta-CNN6 | AS | **0.769** | 0.39 | 0.58 |
| Multi-CNN6 [ours] | AS | 0.73 | **0.4537** | **0.592** |

## 5. Conclusion

In current research, our proposed Multi-breath model, based on the CNN6 architecture, achieved a score of 59.2% on the official split test set of ICBHI 2017, surpassing previously proposed methods based on lightweight models. Multi-breath integrates multi-label training, associating each respiratory sound with a set of category labels represented by a sparse binary label vector, addressing the issue of ambiguous labels that could blur classification boundaries and degrade performance. The categories are defined by the presence of crackle and wheeze features in a respiratory cycle. By replacing the fully connected layer with a multi-head class-specific attention mechanism, we aim to capture spatial relations and important features on the spectrogram over time and dimensions, thereby improving accuracy and reducing the impact of data imbalance.

There is still room for improvement in our model; in the future, we plan to employ consistency learning or knowledge distillation techniques to transfer the feature extraction capabilities of transformers to our lightweight model design. Additionally, we will explore different methods combined with learnable spectrogram front-ends to achieve a more comprehensive analysis. Finally, we hope to enhance the performance of our multi-head attention classifier by referring to the multi-head self-attention approach of transformers. There are limitations to this study due to the use of open-source datasets, where the diversity and age of patients might cause different data distributions. In the future, we will further test our proposed methods on different medical sound classification tasks to verify the generalizability of our model design.